\documentclass[12pt]{article}
\newcommand\eq[1] {(\ref{#1})}

\newcommand\labfig[1] {\label{fig:#1}}

\newcommand{\bfm}[1]{\mbox{\boldmath ${#1}$}}
\newcommand{\nonum}{\nonumber \\}
\newcommand{\beqa}{\begin{eqnarray}}
\newcommand{\eeqa}[1]{\label{#1}\end{eqnarray}}
\newcommand{\beq}{\begin{equation}}
\newcommand{\eeq}[1]{\label{#1}\end{equation}}
\newcommand{\Grad}{\nabla}
\newcommand{\Div}{\nabla \cdot}

\newcommand{\Md}{\partial}

\newcommand{\Ga}{\alpha}
\newcommand{\Gb}{\beta}
\newcommand{\Gd}{\delta}

\newcommand{\Gs}{\sigma}

\newcommand{\Go}{\omega}

\newcommand{\BGve}{\bfm\varepsilon}

\newcommand{\BGm}{\bfm\mu}

\newcommand{\BGr}{\bfm\rho}

\newcommand{\BGs}{\bfm\sigma}

\newcommand{\BCC}{{\bfm{\cal C}}}

\def\Bn{{\bf n}}

\def\Bp{{\bf p}}
\def\Bq{{\bf q}}

\def\Bs{{\bf s}}
\def\Bt{{\bf t}}
\def\Bu{{\bf u}}
\def\Bv{{\bf v}}
\def\Bw{{\bf w}}
\def\Bx{{\bf x}}

\def\BC{{\bf C}}
\def\BD{{\bf D}}
\def\BE{{\bf E}}
\def\BF{{\bf F}}

\def\BI{{\bf I}}

\def\BS{{\bf S}}

\def \ba {\begin{array}}
\def \ea {\end{array}}
\newtheorem {Thm} {Theorem} [section]

\newtheorem {Adef} [Thm] {Definition}

\newtheorem {Arem} [Thm] {Remark}

\newtheorem {Aexa} [Thm] {Example}

\newtheorem {Anot} [Thm] {Notation}

\def \refe #1.{(\ref{#1})}
\def \reff #1.{figure~\ref{#1}}
\def \refs #1.{section~\ref{#1}}
\def \refss #1.{subsection~\ref{#1}}
\def \refD #1.{Definition~\ref{#1}}
\def \refT #1.{Theorem~\ref{#1}}
\def \refL #1.{Lemma~\ref{#1}}
\def \refC #1.{Corollary~\ref{#1}}
\def \refP #1.{Proposition~\ref{#1}}
\def \refR #1.{Remark~\ref{#1}}
\def \refE #1.{Example~\ref{#1}}
\def \refN #1.{Notation~\ref{#1}}

\usepackage{graphics}
\usepackage{amssymb}
\begin{document}
\vspace{-1in}
\title{New metamaterials with macroscopic behavior outside that of continuum elastodynamics}
\author{Graeme W. Milton\\
\small{Department of Mathematics, University of Utah, Salt Lake City UT 84112, USA}}
\date{}
\maketitle
\begin{abstract}
Metamaterials are constructed such that, for a narrow range of frequencies, the momentum density 
depends on the local displacement gradient, and the stress depends on the local velocity.
In these models the momentum density generally depends not only on the strain, 
but also on the local rotation, and the stress is generally not symmetric. A  variant is
constructed for which, at a fixed frequency, the momentum density is independent of the local rotation (but
still depends on the strain) and the stress is symmetric (but still
depends on the velocity). Generalizations of these metamaterials may be useful in
the design of elastic cloaking devices.  
\vskip2mm

\end{abstract}

%%%%%%%%%%%%%%%%%%%%%%%%%%%%%%%%%%%%%%%%%%%%%%%%%%%%%%%%%%%%%%%%%%%%%%%%%
\section{Introduction}
%%%%%%%%%%%%%%%%%%%%%%%%%%%%%%%%%%%%%%%%%%%%%%%%%%%%%%%%%%%%%%%%%%%%%%%%%%
\setcounter{equation}{0}
The possibility of cloaking objects to make them invisible has attracted 
substantial recent attention.  Al\'u and Engheta 
(\cite{Alu:2005:ATP}, \cite{Alu:2007:PMT})
following earlier work of Kerker \cite{Kerker:1975:IB},
found that the scattering from an object could be substantially reduced
by surrounding it by a plasmonic or metamaterial coating. 
Milton and Nicorovici \cite{Milton:2006:CEA}, 
expanding on the work of \cite{Nicorovici:1994:ODP} and
\cite{Milton:2005:PSQ}, proved
that the dipole moments of a polarizable point dipole or clusters of 
polarizable line dipoles would vanish when placed within a cloaking 
region surrounding a superlens, and this has been confirmed numerically
\cite{Nicorovici:2007:OCT}. Bruno And Lintner \cite{Bruno:2007:SCS} show that only partial 
cloaking occurs when the polarizable object is not discrete. 
Ramm \cite{Ramm:1996:MTR} and Miller \cite{Miller:2007:PC}
found that cloaking could be achieved by sensing and manipulating
the fields near the boundary of the object. 

Perhaps the greatest interest has been generated by transformation
based cloaking. This type of cloaking was first discovered by Greenleaf
Lassas and Uhlmann (\cite{Greenleaf:2003:ACC}, \cite{Greenleaf:2003:NCI})
in the context of electrical conductivity. Their idea was to
apply the well known fact (see, for example, \cite{Kohn:1984:IUC}) 
that the equations of electrical conductivity
retain their form under coordinate transformations, using a singular transformation
which mapped a point to a sphere (the cloaking region), and which was the 
identity outside a larger sphere. Perturbing the conductivity in the cloaking
region corresponds to changing the conductivity at a point in the equivalent
problem, and this has no effect on the fields outside the larger sphere.
This type of cloaking was extended to the full time harmonic Maxwell equations by 
Pendry, Schurig and Smith \cite{Pendry:2006:CEM}, and at the same time Leonhardt 
\cite{Leonhardt:2006:OCM} discovered a different
transformation based, two-dimensional, geometric optics or high frequency acoustic
cloaking scheme which only utilized
isotropic materials. The former cloaking has been verified numerically
(\cite{Schurig:2006:CMP}; \cite{Cummer:2006:FWS}; \cite{Zolla:2007:EAC}),
placed on a firm theoretical foundation 
(\cite{Greenleaf:2007:FWI}),
and experimentally substantiated
in the microwave regime in an approximate way 
in two-dimensions by Schurig, Mock, Justice, Cummer, Pendry, Starr, and Smith
\cite{Schurig:2006:MEC} using Schelkunoff and Friis's idea 
(\cite{Schelkunoff:1952:ATP}) of utilizing 
metamaterials with split ring resonators to achieve artificial magnetism. 
Cai, Chettiar, Kildishev, and Shalaev \cite{Cai:2007:OCM}
proposed another design which may experimentally achieve approximate two-dimensional
cloaking in the visible regime.

Milton, Briane and Willis \cite{Milton:2006:CEM} found 
that applying transformation based cloaking to continuum elastodynamics 
would require new materials with very unusual properties. This is because the
continuum elastodynamic equations do not retain their form under coordinate
transformations, but rather take the form of the equations that 
Willis (\cite{Willis:1981:OPC}, \cite{Willis:1981:VPDP}, \cite{Willis:1997:DC})
found described the ensemble averaged behavior of composite materials. One interesting 
feature is that the density at a given frequency is required to be matrix valued. Willis
\cite{Willis:1985:NID} had proved that his density operator had this property. It also follows 
from the work of Movchan and Guenneau \cite{Movchan:2004:SRR} 
and \'Avila, Griso, and Miara \cite{Avila:2005:BPI}, that there exist high contrast 
materials with a local anisotropic effective density. Simple models with anisotropic
and possibly complex density were obtained by Milton and Willis \cite{Milton:2007:MNS}. These 
microstructures generalized a construction of Sheng, Zhang, Liu, and Chan \cite{Sheng:2003:LRS}
and Liu, Chan, and Sheng \cite{Liu:2005:AMP}, who established that the density can be 
negative over a range of frequencies. The same conclusion was
reached and rigorously proved by \'Avila, Griso, and Miara  \cite{Avila:2005:BPI}. A comparison
of Maxwell's equations, which can be written in the form 
\beq \frac{\Md}{\Md x_i}\left( C_{ijk\ell}\frac{\Md E_\ell}{\Md x_k}\right)
=\{\Go^2\BGve\BE\}_j,
\eeq{1.1}
where 
\beq C_{ijk\ell}=e_{ijm}e_{k\ell n}\{\BGm^{-1}\}_{mn},
\eeq{1.2}
[in which $\BE$ is the electric field, $\BGve$ the electric permittivity tensor, 
$\BGm$ the magnetic permeability tensor, and $e_{ijm}=1$ (-1) if $ijm$
is an even (odd) permutation of 123 and is zero otherwise] with
the equations of continuum elastodynamics
\beq \frac{\Md}{\Md x_i}\left( C_{ijk\ell}\frac{\Md u_\ell}{\Md x_k}\right)
=-\{\Go^2\BGr\Bu\}_j,
\eeq{1.3}
[in which $\Bu$ is the displacement field, $\BGr$ is the density, and
$\BC$ is now the elasticity tensor] suggests that $\BGr$ as a function
of the frequency $\Go$ could share many of the same properties as
$\BGve(\Go)$. This has been established, and in fact for every function
$\BGr(\Go)$ satisfying these properties one can construct a model
which has approximately that function as its effective density as a function
of frequency (\cite{Milton:2007:MNS}). Cummer and Schurig \cite{Cummer:2007:PAC}
investigated transformation based acoustic cloaking in two-dimensions 
and, due to a mathematical equivalence with the electromagnetic problem,
found that it could be achieved provided one could construct the
necessary materials with anisotropic density.

Anisotropic density is not the only new property needed to achieve transformation
based elasticity cloaking. One needs materials where the constitutive law,
like that in the Willis equations, couples the stress with not only with
the strain but also with the velocity, and couples the momentum density not only
with the velocity but also with the displacement gradient through the strain.
Also, unlike in the Willis equations, one wants this dependence to be local in space
and to apply to a single microgeometry rather than to an ensemble average of
microgeometries. Here we show that such unusual behavior can be realized
in a model such that, for a narrow range of frequencies, the associated 
waves have wavelength much larger than the microstructure, and the momentum density 
depends on the displacement gradient, and the stress depends on the velocity.
Curiously the momentum density depends not only on the strain, 
but also on the local rotation, and the stress is not symmetric.
We also construct a variant of the model for which (at a fixed
frequency) the momentum density is independent of the local rotation (but
still depends on the strain) and the stress is symmetric (but still
depends on the velocity). Both models rely heavily on the discovery
Sheng, Zhang, Liu, and Chan \cite{Sheng:2003:LRS} that
one can design structures which, at a fixed frequency, respond 
as if they have negative mass. Following their ideas, and the
simplified constructions of Milton and Willis \cite{Milton:2007:MNS} which
incorporate these ideas, figure 1 illustrates
a structure with negative mass. In the models considered in this
paper both positive and
negative masses are idealized as point masses.

\begin{figure}
\vspace{2in}
\hspace{1.0in}
{\resizebox{2.0in}{1.0in}
{\includegraphics[0in,0in][4in,2in]{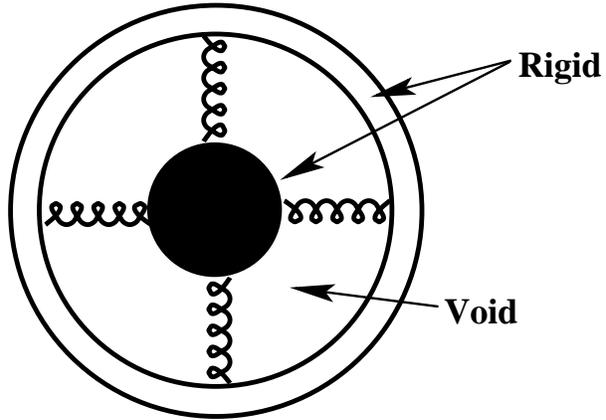}}}
\vspace{0.1in}
\caption{This structure responds as if it had negative mass
to oscillations at a fixed frequency above resonance, such that the spherical core oscillates out
of phase with the rigid, but light, surrounding shell. The springs have the same spring 
constant to ensure that the effective mass is isotropic.}
\labfig{-1}
\end{figure}

This work on metamaterials with macroscopic behavior outside that of continuum elastodynamics, even 
though they are governed by continuum elastodynamics at the microscale, is preceeded
by  work on metamaterials with macroscopic non-Ohmic, possibly non-local,  conducting behavior, 
even though they conform to Ohm's
law at the microscale, 
(\cite{Khruslov:1978:ABS}; \cite{Briane:1998:HSW}; \cite{Briane:1998:HTR}; \cite{Briane:2002:HNU};
\cite{Camar:2002:CSD}; \cite{Cherednichenko:2006:NLH})
by work on metamaterials with non-Maxwellian macroscopic electromagnetic behavior, even though they 
conform to Maxwell's equations at the microscale (\cite{Shin:2007:TDE}), and by work on metamaterials with
a macroscopic higher order gradient or non-local elastic response even though they are governed by usual linear 
elasticity equations at the microscale 
(\cite{Bouchitte:2002:HSE}; \cite{Alibert:2003:TMB}; \cite{Camar:2003:DCS}).

For simplicity we consider a two dimensional model, although it can
be generalized to three dimensions. The model is illustrated in
figure 2. 
\begin{figure}
\vspace{2in}
\hspace{1.0in}
{\resizebox{2.0in}{1.0in}
{\includegraphics[0in,0in][16in,8in]{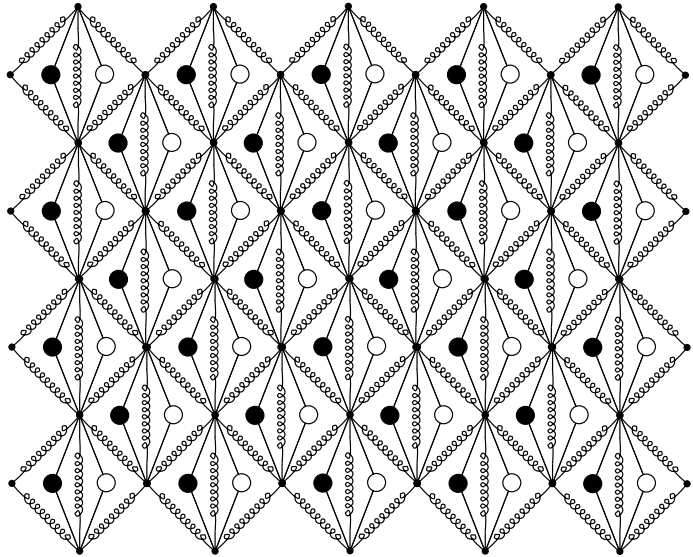}}}
\vspace{0.1in}
\caption{The model with strange elastic behavior. The large solid black circles
have positive mass, while the neighbouring large white circles have 
negative mass at the given frequency.
They are connected to the spring network by rods of fixed length, which alternatively
can be regarded as springs with infinite spring constants.}
\labfig{0}
\end{figure}
%%%%%%%%%%%%%%%%%%%%%%%%%%%%%%%%%%%%%%%%%%%%%%%%%%%%%%%%%%%%%%%%%%%%%%%%%%%%%%%%%%%%%%%%%%%%%%%%%%%%%%%%%%%%%
\section{The momentum density in the model}
%%%%%%%%%%%%%%%%%%%%%%%%%%%%%%%%%%%%%%%%%%%%%%%%%%%%%%%%%%%%%%%%%%%%%%%%%%%%%%%%%%%%%%%%%%%%%%%%%%%%%%%%%%%%%
\setcounter{equation}{0}
Figure 3 shows the unit cell in the model. The masses are approximated as
point masses, and the springs may be taken to have all the same spring
constant $hK$, which scales in proportion to the size of the unit cell.
This ensures that the spring network, by itself, responds as a elastic
material in the limit $h\to 0$.
\begin{figure}
\vspace{2.5in}
\hspace{1.0in}
{\resizebox{2.0in}{1.0in}
{\includegraphics[0in,0in][4in,2in]{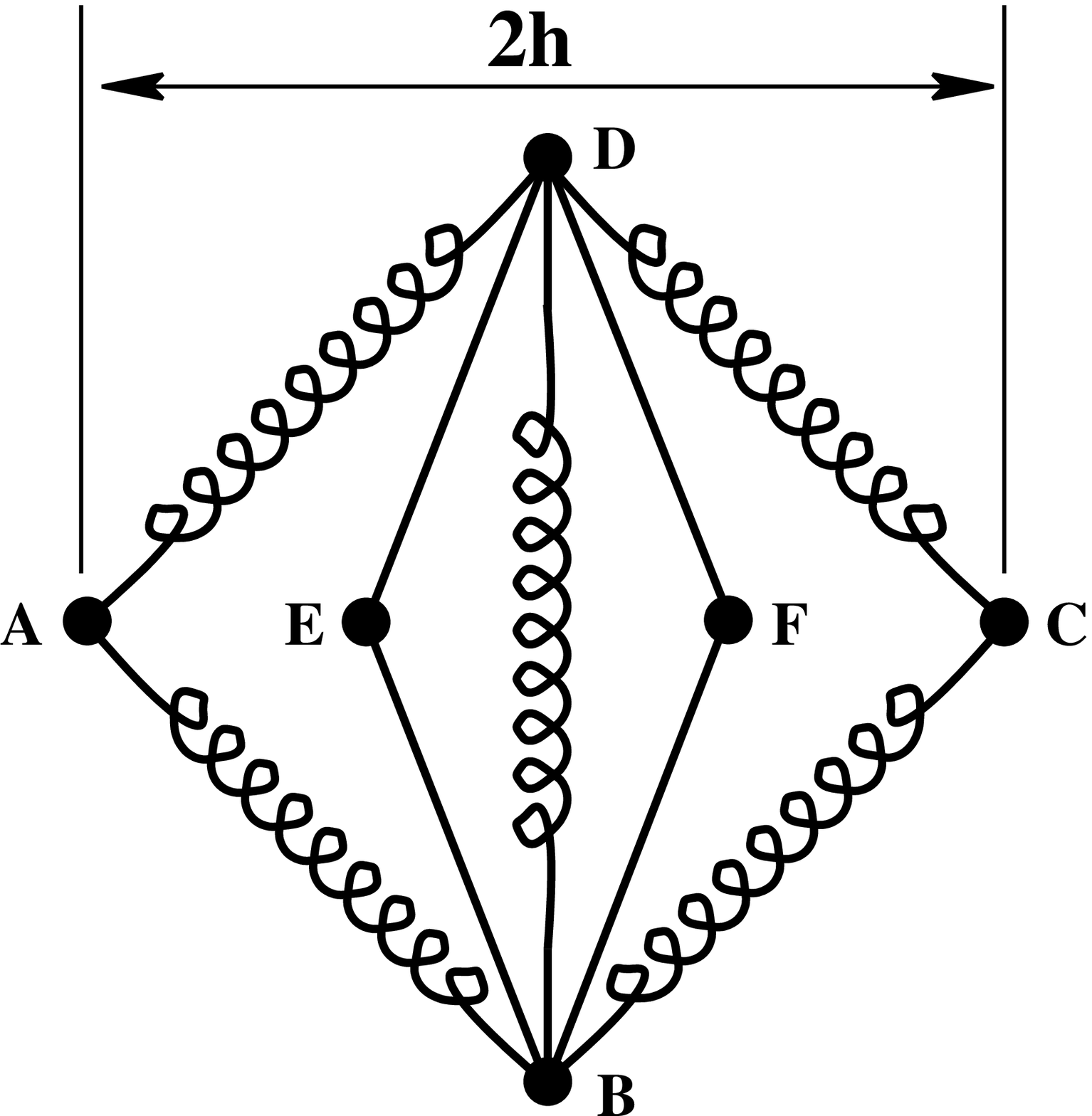}}}
\vspace{0.1in}
\caption{The unit cell of the model.}
\labfig{1}
\end{figure}

Without loss of generality let us place the origin at the center $\Bx_0$
of the unit cell under consideration. Then when the material is a rest the
points $A,B,C,D,E$ and $F$ in figure 3 are at positions
\beqa \Bx_A& = & (-h,0),\quad \Bx_B=(0,-h),\quad \Bx_C=(h,0),\quad \Bx_D=(0,h) \nonum
\quad \Bx_E& = & (-ch,0),\quad \Bx_F=(ch,0),
\eeqa{2.1}
where $c$ is a parameter between 0 and 1 which controls the inclination
of the rods joining $E$ and $F$ to $B$ and $D$: these
rods have length $h\sqrt{1+c^2}$. In the limit
$h\to 0$ we assume the (infinitesimal) physical displacements at the 
points $\Bx_A$, $\Bx_B$, $\Bx_C$ and $\Bx_D$ derive from some smooth 
complex valued displacement field $\Bu(\Bx)$, and in particular
\beqa 
\Bu_A & = & \Re e[\Bu(\Bx_A)e^{-i\omega t}] \approx  \Re e[(\Bu_0-h\Bq)e^{-i\omega t}],
\nonum
 \Bu_B & = & \Re e[\Bu(\Bx_B)e^{-i\omega t}] \approx  \Re e[(\Bu_0-h\Bw)e^{-i\omega t}],
\nonum
 \Bu_C & = & \Re e[\Bu(\Bx_A)e^{-i\omega t}] \approx  \Re e[(\Bu_0+h\Bq)e^{-i\omega t}],
\nonum
 \Bu_D & = & \Re e[\Bu(\Bx_D)e^{-i\omega t}] \approx \Re e[(\Bu_0+h\Bw)e^{-i\omega t}],
\nonum
\eeqa{2.2}
where 
\beq \Bu_0=\Bu(\Bx_0),\quad \Bq=\frac{\Md\Bu}{\Md x_1}\Bigg{|}_{\Bx=\Bx_0}, \quad
\Bw=\frac{\Md\Bu}{\Md x_2}\Bigg{|}_{\Bx=\Bx_0},
\eeq{2.3}
and $\Bx_0=0$ because the origin is at the center of the unit cell.

At the macroscopic level we choose not keep track of the displacements $\Bu_E$ and
$\Bu_F$. These are internal hidden variables which however can be recovered
from $\Bu_B$ and $\Bu_D$. They have the form  
\beqa  \Bu_E & \approx & \Re e[(\Bu_0+h\Bs)e^{-i\omega t}],
\nonum
 \Bu_F & \approx & \Re e[(\Bu_0-h\Bs)e^{-i\omega t}],
\eeqa{2.4}
where the complex vector $\Bs$ remains to be determined. 
Since there is a rigid rod connecting the points $D$ and $E$ we have,
in the limit $h\to 0$, the constraint that $\Bw-\Bs$ must be perpendicular to
the line joining $D$ and $E$, i.e.
\beq 0=(\Bw-\Bs)\cdot(c,1)=c(w_1-s_1)+(w_2-s_2). \eeq{2.5}
Similarly since there is a rigid rod connecting the points $B$ and $E$ we have,
in the limit $h\to 0$, the constraint
\beq 0=(\Bw+\Bs)\cdot(-c,1)=-c(w_1+s_1)+(w_2+s_2). \eeq{2.6}
These equations have the solution
\beq s_1=w_2/c=\frac{1}{c}\frac{\Md u_2}{\Md x_2},
\quad s_2=cw_1=c\frac{\Md u_1}{\Md x_2}.
\eeq{2.10}
One can easily see that if $c$ is replaced by $-c$ then $\Bs$ gets
replaced by $-\Bs$, which justifies the formula \eq{2.4} for $\Bu_F$.

Now suppose the masses at the points $E$ and $F$, which we call
a mass pair, are respectively chosen 
to have the form
\beq m_E=hm, \quad m_F=-hm+\Gd h^2, \eeq{2.11}
where $m$ is a positive real constant, and $\Gd$ is a possibly complex constant
with a non-negative imaginary part. Then at any time $t$ the physical 
momentum in the unit cell is, to second order in $h$,
\beq \Re e\{-i\omega[m_E(\Bu_0+h\Bs)+m_F(\Bu_0-h\Bs)]e^{-i\omega t}\}
\approx  h^2\Re e\{-i\omega[2m\Bs+\Gd\Bu_0]e^{-i\omega t}\}.
\eeq{2.12}
Since the area of the unit cell is $2h^2$ we see that the complex momentum 
density is
\beq \Bp=-i\omega m\Bs+(\Gd/2)(-i\Go\Bu_0),
\eeq{2.13}
in which $-i\Go\Bu_0$ is the complex velocity of the unit cell, and
$\Bs$ depends on the deformation gradient through \eq{2.10}.
%%%%%%%%%%%%%%%%%%%%%%%%%%%%%%%%%%%%%%%%%%%%%%%%%%%%%%%%%%%%%%%%%%%%%%%%%%%%%%%%%%%%%%%%%%%%%%%%%%%%%%%%%%%%%
\section{The stress in the model}
%%%%%%%%%%%%%%%%%%%%%%%%%%%%%%%%%%%%%%%%%%%%%%%%%%%%%%%%%%%%%%%%%%%%%%%%%%%%%%%%%%%%%%%%%%%%%%%%%%%%%%%%%%%%%
\setcounter{equation}{0}
The acceleration of the material will also generate stress. To see 
this, note that to leading order in $h$, the masses at $E$ and $F$ must be 
accelerated by complex forces $\BF=-\Go^2mh\Bu$ and $-\BF$ acting on
the respective masses (the physical forces
are obtained by multiplying these by $e^{-i\omega t}$ and taking 
the real part). Let $\BF_{ED}$ and $\BF_{FD}$ be the forces which the rods 
$ED$ and $FD$ exert on the vertex $D$ and let $\BF_{EB}$ and $\BF_{FB}$ 
be the forces which the rods $EB$ and $FB$ exert on the vertex $B$. Balance
of forces at $E$ and $F$ requires that
\beq \BF_{ED}+\BF_{EB}=-\BF,\quad \BF_{FD}+\BF_{FB}=\BF.
\eeq{3.14}
Since these forces are aligned with their respective rods we have
\beq \BF_{ED}=-\BF_{FB}=\Ga(c,1),\quad \BF_{EB}=-\BF_{FD}=\Gb(c,-1),
\eeq{3.15}
where the constants $\Ga$ and $\Gb$ are such that \eq{3.14} is
satisfied:
\beq (\Ga+\Gb)c=-F_1,\quad \Ga-\Gb=-F_2.
\eeq{3.16}
Forgetting for a moment the springs and surrounding material,
to maintain the motions a force 
\beq h\Bt=-\BF_{ED}-\BF_{FD}
=-\Ga(c,1)+\Gb(c,-1)=(cF_2,F_1/c)=-\Go^2mh(cu_2,u_1/c)
\eeq{3.17}
needs to act on the vertex $D$ to balance the forces which the rods 
$ED$ and $FD$ exert on this vertex. Similarly a force $-h\Bt$ 
needs to act on the vertex $B$ to balance the forces which the rods 
$EB$ and $FB$ exert on this vertex. On the other hand, no additional
forces need to be exerted on the vertices $A$ and $C$. 

     Now consider a rectangular sample (with dimensions, say of order
$\sqrt{h}$, small compared to the wavelength) as in figure 2
with (complex) forces acting on the boundary 
vertices to maintain the motions. These forces will have two components:
an elastic (or possibly viscoelastic) component to compensate 
for the (complex) stress $\BGs_E$ caused by the (complex) strain
in the spring network, and an inertial component to compensate for the
inertial stress $\BGs_I$ caused by the acceleration $-\Go^2\Bu$ of the material.
The compensating inertial component will be zero on the left and right sides of the
network. At the top 5 vertices in figure 2 the compensating inertial
component will be approximately $2h\Bt$ at each vertex. The extra factor of $2$
arises because of the extra load that the springs at the top boundary carry
to support the inertial component of the forces at the 4 vertices immediately
below the top. (At the other interior nodes these forces are balanced). 
Thus the compensating inertial component per unit length,
is $\Bt$ on the top, $-\Bt$ on the bottom, and zero on the sides. By the
definition of stress this should be equated with $\BGs_I\Bn$, where
$\Bn$ is the unit normal to the boundary. Thus we deduce that   
\beq \BGs_I=\pmatrix{0 & t_1 \cr 0 & t_2\cr}
=\pmatrix{0 & -\Go^2mcu_2 \cr 0 & -\Go^2mu_1/c \cr}.
\eeq{3.18}
Of course, in the limit $h\to 0$, the stress in the spring network 
(excluding the masses and connecting rods) 
will be governed by a normal relation
\beq \BGs_E=\BCC[\Grad\Bu+(\Grad\Bu)^T],
\eeq{3.19}
where $\BCC$ is the elasticity tensor of the network.
      
Our choice to define the macroscopic stress through the forces at the
vertices of the unit cell (rather than as a volume average of the
microscopic stress field, which would result in a symmetric stress)
makes good physical sense and is consistent with our choice to
define the macroscopic displacement field through the values of the
displacement at the vertices of the unit cell. These choices
bear some resemblance to the way Pendry, Holden, Robbins and Willis \cite{Pendry:1999:MCE} 
define macroscopic electromagnetic
fields through field values at the boundary of the unit cell. In defining the stress
in this way it is 
important that the unit cell be chosen so that each mass pair (of positive 
and negative masses) is not split by the boundary of the unit cell:
each mass pair and their four connecting rods is regarded as an indivisible unit in the same way that
in defining the electric polarization field
one would not choose to place the boundary in a dielectric material
so as to split the positive and negative charges of the constituent
atoms.

%%%%%%%%%%%%%%%%%%%%%%%%%%%%%%%%%%%%%%%%%%%%%%%%%%%%%%%%%%%%%%%%%%%%%%%%%%%%%%%%%%%%%%%%%%%%%%%%%%%%%%%%%%%%%
\section{Summary}
%%%%%%%%%%%%%%%%%%%%%%%%%%%%%%%%%%%%%%%%%%%%%%%%%%%%%%%%%%%%%%%%%%%%%%%%%%%%%%%%%%%%%%%%%%%%%%%%%%%%%%%%%%%%%
\setcounter{equation}{0}
In the limit $h\to 0$ the constitutive law takes the form
\beq \pmatrix{\BGs \cr \Bp}=
\pmatrix{\BCC & \BS \cr \BD & \BGr}\pmatrix{\Grad\Bu \cr \Bv},
\eeq{4.1}
where $\BGs=\BGs_E+\BGs_I$ is the total stress, $\Bv=-i\Go\Bu$ is the 
velocity, and $\BGr=(\Gd/2)\BI$ is
the density. In the basis where $\BGs$, $\Bp$, $\Grad\Bu$ and $\Bv$
are represented by the vectors
\beq \BGs=\pmatrix{\Gs_{11} \cr \Gs_{21} \cr \Gs_{12} \cr \Gs_{22}},\quad
\Bp=\pmatrix{p_1\cr p_2},\quad
\Grad\Bu=\pmatrix{\Md u_1/\Md x_1 \cr \Md u_2/\Md x_1 
              \cr \Md u_1/\Md x_2 \cr \Md u_2/\Md x_2},\quad
\Bv=\pmatrix{v_1\cr v_2},
\eeq{4.2}
the third-order tensors $\BS$ and $\BD$, from \eq{2.10}, \eq{2.13}
and \eq{3.18}, are represented by the matrices 
\beq \BS=\pmatrix{ 0 & 0 \cr
                   0 & 0 \cr
                   0 & -i\Go mc \cr
                   -i\Go m/c & 0 \cr}, \quad
\BD=\pmatrix{0 & 0 & 0 & -i\Go m/c \cr
             0 & 0 & -i\Go m c & 0 \cr}.
\eeq{4.3}
Thus the matrix entering the constitutive law \eq{4.1} is symmetric. Since
the constitutive law and the equation of motion are asymptotically independent
of $h$, the waves associated with their solutions will have wavelength much
larger that the microstructure. In this circumstance the usual continuum
elastodynamic equations normally apply. However this model shows that this
is not always the case.

The model has some unusual features. Although the net amount of 
mass in a region of unit area is independent of $h$, the total
amount of positive (or negative) mass in a region of unit area
scales like $1/h$. However in any physical model, as usual,
one never in fact takes the limit $h\to 0$, but instead one sets the
microstructure as small as required to get a reasonable approximation
to the limiting behavior for a desired set of applied fields. Also the
amount of positive (or negative) mass in the unit cell only
represents the amount of apparent mass, not gravitational mass, which
could be quite different if the apparent mass arises from substructures
which are close to resonance. Of course
the analysis given here needs some rigorous justification to check, for 
example, that the displacements can be approximated by a smooth field
$\Bu(\Bx)$ in the limit $h\to 0$. Also from a practical point of view 
it needs to be checked that the behavior is stable to sufficiently
small variations in the masses or spring constants from cell to cell.
\begin{figure}
\vspace{2.5in}
\hspace{1.0in}
{\resizebox{2.0in}{1.0in}
{\includegraphics[0in,0in][4in,2in]{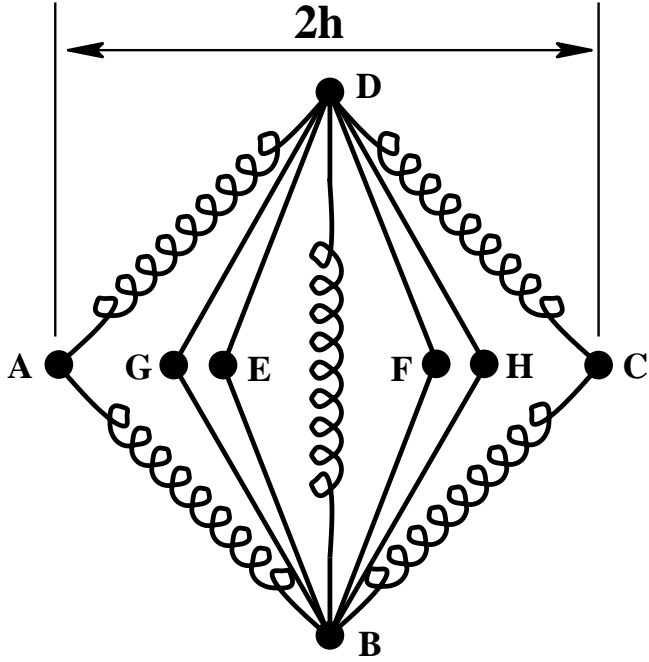}}}
\vspace{0.1in}
\caption{The unit cell of a generalized model.}
\labfig{3}
\end{figure}

The model can easily be generalized. One simple generalization has 
additional masses and connecting rods as illustrated in figure 4,
with masses 
\beq m_G=-hm'+\Gd h^2, \quad m_H=hm', \eeq{4.4}
at the points $G$ and $H$ which, when the material is at rest,
are located at
\beq  \Bx_G =  (-c'h,0),\quad \Bx_H=(c'h,0),
\eeq{4.5}
where $c'$ and $m'$ are positive constants.
By superposition the constitutive law will take the form
\eq{4.1} with a density $\BGr=\Gd\BI$ and with the third-order tensors 
$\BS$ and $\BD$ represented by the matrices 
\beqa \BS& = & \pmatrix{ 0 & 0 \cr
                   0 & 0 \cr
                   0 & i\Go(m'c'-mc) \cr
                    i\Go(m'/c'- m/c) & 0 \cr}, \nonum
\BD& = &\pmatrix{0 & 0 & 0 & i\Go(m'/c'- m/c) \cr
             0 & 0 & i\Go(m'c'-m c) & 0 \cr}.
\eeqa{4.6}
In particular if, at a given frequency, $m'c'=mc$ we obtain a model
in which the momentum does not depend on the local
rotation and the stress is symmetric. With this condition satisfied
one has the relations
\beq \BGs=\BCC\Grad\Bu+\BS'\Bu,\quad\quad\Div\BGs=-i\Go\Bp=\BD'\Grad\Bu-\Go^2\BGr\Bu,
\eeq{4.7}
where
\beq \BS'=-i\Go\BS,\quad\quad \BD'=-i\Go\BD,
\eeq{4.8}
and the only non-zero Cartesian elements of these two tensors are
from \eq{4.1}, \eq{4.2} and \eq{4.6}
\beq S'_{221}=D'_{122}=\Go^2(m'/c'- m/c).
\eeq{4.9}
The general form of these tensors $\BS'$ and $\BD'$ corresponds with
that required for elasticity cloaking: they are real and satisfy the
identities $S'_{ijk}=S'_{jik}=D'_{kij}$. They may serve as a good
starting point in the quest to find materials with a suitable
combination of properties $\BCC$, $\BS'$, $\BD'$ and $\BGr$ 
needed for elasticity cloaking. However, quite apart from this,
the novel properties of this new class of metamaterials
may have other important applications.

In the design of these models perhaps the key feature is that
in the narrow frequency range under consideration
the amount of negative mass in the unit cell (due to hidden internal masses
moving out of phase with the motion) almost balances the amount
of positive mass in the unit cell. This ensures that the momentum
density approaches a constant as the cell size approaches zero,
while keeping non-trivial the stresses caused by the moving masses in the unit
cell. It will be interesting to see if other models with this key feature also have
a local constitutive law of the form \eq{4.1}, with non-zero
coupling terms $\BS$ and $\BD$.

\section*{Acknowledgements}

Graeme Milton is grateful to John Willis for
helpful suggestions which improved the manuscript
and for support from the 
National Science Foundation through grant DMS-0411035.

\bibliography{/u/ma/milton/tcbook,/u/ma/milton/newref}

\ifx \bblindex \undefined \def \bblindex #1{} \fi\ifx \bblindex \undefined \def
  \bblindex #1{} \fi
\begin{thebibliography}{10}

\bibitem{Alu:2005:ATP}
A. Al\'u and N. Engheta, ``Achieving transparency with plasmonic and
  metamaterial coatings,'' Physical Review E (Statistical Physics, Plasmas,
  Fluids, and Related Interdisciplinary Topics) {\bf 72,} 0166623 (2005).

\bibitem{Alu:2007:PMT}
A. Al\'u and N. Engheta, ``Plasmonic materials in transparency and cloaking
  problems: mechanism, robustness, and physical insights,'' Optics Express {\bf
  15,} 3318--3332 (2007).

\bibitem{Kerker:1975:IB}
M. Kerker, ``Invisible bodies,'' Journal of the Optical Society of America {\bf
  65,} 376--379 (1975).

\bibitem{Milton:2006:CEA}
G.~W. Milton and N.-A.~P. Nicorovici, ``On the cloaking effects associated with
  anomalous localized resonance,'' Proceedings of the Royal Society of London.
  Series A, Mathematical and Physical Sciences {\bf 462,} 3027--3059 (2006),
  published online May 3rd: doi:10.1098/rspa.2006.1715.

\bibitem{Nicorovici:1994:ODP}
N.~A. Nicorovici, R.~C. McPhedran, and G.~W. Milton, ``Optical and dielectric
  properties of partially resonant composites,'' Physical Review B (Solid
  State) {\bf 49,} 8479--8482 (1994).

\bibitem{Milton:2005:PSQ}
G.~W. Milton, N.-A.~P. Nicorovici, R.~C. McPhedran, and V.~A. Podolskiy, ``A
  proof of superlensing in the quasistatic regime, and limitations of
  superlenses in this regime due to anomalous localized resonance,''
  Proceedings of the Royal Society of London. Series A, Mathematical and
  Physical Sciences {\bf 461,} 3999--4034 (2005).

\bibitem{Nicorovici:2007:OCT}
N.-A.~P. Nicorovici, G.~W. Milton, R.~C. McPhedran, and L.~C. Botten,
  ``Quasistatic cloaking of two-dimensional polarizable discrete systems by
  anomalous resonance,'' Optics Express {\bf 15,} 6314--6323 (2007).

\bibitem{Bruno:2007:SCS}
O.~P. Bruno and S. Lintner, ``Superlens-cloaking of small dielectric bodies in
  the quasistatic regime,''   (2007), submitted.

\bibitem{Ramm:1996:MTR}
A.~G. Ramm, ``Minimization of the total radiation from an obstacle by a control
  function on a part of its boundary,'' Journal of Inverse and Ill-Posed
  Problems {\bf 4,} 531--534 (1996).

\bibitem{Miller:2007:PC}
D.~A.~B. Miller, ``On perfect cloaking,'' Optics Express {\bf 14,} 12457--12466
  (2006).

\bibitem{Greenleaf:2003:ACC}
A. Greenleaf, M. Lassas, and G. Uhlmann, ``Anisotropic conductivities that
  cannot be detected by {EIT},'' Physiological Measurement {\bf 24,} 413--419
  (2003).

\bibitem{Greenleaf:2003:NCI}
A. Greenleaf, M. Lassas, and G. Uhlmann, ``On non-uniqueness for {Calder\'on's}
  inverse problem,'' Mathematical Research Letters {\bf 10,} 685--693 (2003).

\bibitem{Kohn:1984:IUC}
R.~V. Kohn and M. Vogelius, ``Inverse Problems,'' In {\em Proceedings of the
  Symposium in Applied Mathematics of the American Mathematical Society and the
  Society for Industrial and Applied Mathematics, New York, April 12-13, 1983},
  D.~W. McLaughlin, ed.,\ SIAM AMS Proceedings\ {\bf 14,} 113--123  (American
  Mathematical Society, Providence, Rhode Island, 1984).

\bibitem{Pendry:2006:CEM}
J.~B. Pendry, D. Schurig, and D.~R. Smith, ``Controlling electromagnetic
  fields,'' Science {\bf 312,} 1780--1782 (2006).

\bibitem{Leonhardt:2006:OCM}
U. Leonhardt, ``Optical conformal mapping,'' Science {\bf 312,} 1777--1780
  (2006).

\bibitem{Schurig:2006:CMP}
D. Schurig, J.~B. Pendry, and D.~R. Smith, ``Calculation of material properties
  and ray tracing in transformation media,'' Optics Express {\bf 14,}
  9794--9804 (2006).

\bibitem{Cummer:2006:FWS}
S.~A. Cummer, B.-I. Popa, D. Schurig, D.~R. Smith, and J. Pendry, ``Full-wave
  simulation of electromagnetic cloaking structures,'' Physical Review E
  (Statistical Physics, Plasmas, Fluids, and Related Interdisciplinary Topics)
  {\bf 74,} 036621 (2006).

\bibitem{Zolla:2007:EAC}
F. Zolla, S. Guenneau, A. Nicolet, and J.~B. Pendry, ``Electromagnetic analysis
  of cylindrical invisibility cloaks and the mirage effect,'' Optics Letters
  {\bf 32,} 1069--1071 (2007).

\bibitem{Greenleaf:2007:FWI}
A. Greenleaf, Y. Kurylev, M. Lassas, and G. Uhlmann, ``Full-wave invisibility
  of active devices at all frequencies,'' Communications in Mathematical
  Physics  (2007), to appear, see also arXiv:math/0611185v3.

\bibitem{Schurig:2006:MEC}
D. Schurig, J.~J. Mock, B.~J. Justice, S.~A. Cummer, J.~B. Pendry, A.~F. Starr,
  and D.~R. Smith, ``Metamaterial electromagnetic cloak at microwave
  frequencies,'' Science {\bf 314,} 977--980 (2006).

\bibitem{Schelkunoff:1952:ATP}
S.~A. Schelkunoff and H.~T. Friis,  in {\em Antennas: the theory and practice}
  (John Wiley and Sons, New York~/ London~/ Sydney, Australia, 1952), \ pp.\
  584--585.

\bibitem{Cai:2007:OCM}
W. Cai, U.~K. Chettiar, A.~V. Kildishev, and V.~M. Shalaev, ``Optical cloaking
  with metamaterials,'' Nature Photonics {\bf 1,} 224--226 (2007).

\bibitem{Milton:2006:CEM}
G.~W. Milton, M. Briane, and J.~R. Willis, ``On cloaking for elasticity and
  physical equations with a transformation invariant form,'' New Journal of
  Physics {\bf 8,} 248 (2006).

\bibitem{Willis:1981:OPC}
J.~R. Willis, ``Variational and related methods for the overall properties of
  composites,'' Advances in Applied Mechanics {\bf 21,} 1--78 (1981).

\bibitem{Willis:1981:VPDP}
J.~R. Willis, ``Variational principles for dynamic problems for inhomogeneous
  elastic media,'' Wave Motion {\bf 3,} 1--11 (1981).

\bibitem{Willis:1997:DC}
J.~R. Willis, ``Dynamics of Composites,'' In {\em Continuum Micromechanics,
  CISM Lecture Notes}, \ pp.\ 265--290  (Springer, Wein/New York, 1997).

\bibitem{Willis:1985:NID}
J.~R. Willis, ``The non-local influence of density variations in a composite,''
  International Journal of Solids and Structures {\bf 21,} 805--817 (1985).

\bibitem{Movchan:2004:SRR}
A.~B. Movchan and S. Guenneau, ``Split-ring resonators and localized modes,''
  Physical Review B (Solid State) {\bf 70,} 125116 (2004).

\bibitem{Avila:2005:BPI}
A. \'Avila, G. Griso, and B. Miara, ``Bandes phononiques interdites en
  \'elasticit\'e lin\'earis\'ee,'' Physical Review E (Statistical Physics,
  Plasmas, Fluids, and Related Interdisciplinary Topics) {\bf 340,} 933--938
  (2005).

\bibitem{Milton:2007:MNS}
G.~W. Milton and J.~R. Willis, ``On modifications of {Newton's} second law and
  linear continuum elastodynamics,'' Proceedings of the Royal Society of
  London. Series A, Mathematical and Physical Sciences {\bf 463,} 855--880
  (2007).

\bibitem{Sheng:2003:LRS}
P. Sheng, X.~X. Zhang, Z. Liu, and C.~T. Chan, ``Locally resonant sonic
  materials,'' Physica. B, Condensed Matter {\bf 338,} 201--205 (2003).

\bibitem{Liu:2005:AMP}
Z. Liu, C.~T. Chan, and P. Sheng, ``Analytic model of phononic crystals with
  local resonances,'' Physical Review B (Solid State) {\bf 71,} 014103 (2005).

\bibitem{Cummer:2007:PAC}
S.~A. Cummer and D. Schurig, ``One path to acoustic cloaking,'' New Journal of
  Physics {\bf 9,} 45 (2007).

\bibitem{Khruslov:1978:ABS}
E.~Y. Khruslov, ``Asymptotic behavior of the solutions of the second boundary
  value problem in the case of the refinement of the boundary of the domain,''
  Matematicheskii sbornik {\bf 106,} 604--621 (1978), english translation in
  {\em Math. USSR Sbornik} 35:266--282 (1979).

\bibitem{Briane:1998:HSW}
M. Briane, ``Homogenization in some weakly connected domains,'' Ricerche di
  Matematica (Napoli) {\bf 47,} 51--94 (1998).

\bibitem{Briane:1998:HTR}
M. Briane and L. Mazliak, ``Homogenization of two randomly weakly connected
  materials,'' Portugaliae Mathematica {\bf 55,} 187--207 (1998).

\bibitem{Briane:2002:HNU}
M. Briane, ``Homogenization of non-uniformly bounded operators: critical
  barrier for nonlocal effects,'' Archive for Rational Mechanics and Analysis
  {\bf 164,} 73--101 (2002).

\bibitem{Camar:2002:CSD}
M. Camar-Eddine and P. Seppecher, ``Closure of the set of diffusion functionals
  with respect to the {Mosco}-convergence,'' Mathematical Models and Methods in
  Applied Sciences {\bf 12,} 1153--1176 (2002).

\bibitem{Cherednichenko:2006:NLH}
K.~D. Cherednichenko, V.~P. Smyshlyaev, and V.~V. Zhikov, ``Non-local
  homogenized limits for composite media with highly anisotropic periodic
  fibres,'' Proceedings of the Royal Society of Edinburgh {\bf 136A,} 87--114
  (2006).

\bibitem{Shin:2007:TDE}
J. Shin, J.-T. Shen, and S. Fan, ``Three-dimensional electromagnetic
  metamaterials with non-Maxwellian effective fields,''   (2007),
  arXiv:physics/0703053v1 (2005).

\bibitem{Bouchitte:2002:HSE}
G. Bouchitt\'e and M. Bellieud, ``Homogenization of a soft elastic material
  reinforced by fibers,'' Asymptotic Analysis {\bf 32,} 153--183 (2002).

\bibitem{Alibert:2003:TMB}
J.-J. Alibert, F. dell'Isola, and P. Seppecher, ``Truss modular beams with
  deformation energy depending on higher displacement gradients,'' Mathematics
  and Mechanics of Solids {\bf 8,} 51--74 (2003).

\bibitem{Camar:2003:DCS}
M. Camar-Eddine and P. Seppecher, ``Determination of the closure of the set of
  elasticity functionals,'' Archive for Rational Mechanics and Analysis {\bf
  170,} 211--245 (2003).

\bibitem{Pendry:1999:MCE}
J. Pendry, A.~J. Holden, D.~J. Robbins, and W.~J. Stewart, ``Magnetism from
  conductors and enhanced nonlinear phenomena,'' IEEE Transactions on Microwave
  Theory and Techniques {\bf 47,} 2075--2084 (1999).

\end{thebibliography}

\end{document}